\documentclass[%
 aip,
 amsmath,amssymb,
 reprint,%
]{revtex4-1}

\usepackage{graphicx}
\usepackage{dcolumn}
\usepackage{bm}

\usepackage[utf8]{inputenc}
\usepackage[T1]{fontenc}
\usepackage{mathptmx}
\usepackage{etoolbox}

\makeatletter
\def\@email#1#2{%
 \endgroup
 \patchcmd{\titleblock@produce}
  {\frontmatter@RRAPformat}
  {\frontmatter@RRAPformat{\produce@RRAP{*#1\href{mailto:#2}{#2}}}\frontmatter@RRAPformat}
  {}{}
}%
\makeatother
\begin{document}

\preprint{AIP/123-QED}

\title{Considering a Superposition of Classical Reference Frames}
\author{Elliott Tammaro}
 \affiliation{Department of Physics, Chestnut Hill College, 9601 Germantown Ave., Philadelphia, 19118, PA, USA }
 \author{H. Angle}
 \affiliation{Department of Biology, University of Delaware, 210 South College Ave., Newark, 19716, DE, USA}

\author{E. Mbadu}
\affiliation{Department of Computer Science, Drexel University, 3141 Chestnut St., Philadelphia, 19104, PA, USA}

\date{20 December 2023}
Journal of Mathematical Physics, Vol. 64, Issue 12
DOI: 10.1063/5.0144924

\begin{abstract}
 A ubiquitous feature of quantum mechanical theories is the existence of states of superposition. This is expected to be no different for a quantum gravity theory. Guided by this consideration and others we consider a framework in which classical reference frames may be in superposition relative to one another. Mirroring standard quantum mechanics we introduce a complex-valued wavefunctional, which takes as input the transformations between the coordinates, $\Psi[x(x')]$, with the interpretation that an interaction between the reference frames may select a particular transformation with probability distribution given by the Born rule --- $P[x(x')] = \text{probability distribution functional} \equiv \vert \Psi[x(x')] \vert^2$. The cases of two and three reference frames in superposition are considered explicitly. It is shown that the set of transformations is closed. A rule for transforming wavefunctions from one system to another system in superposition is proposed and consistency with the Schrodinger equation is demonstrated.
\end{abstract}

\maketitle

\section{\label{sec1}Introduction}
 Quantum theories, by virtue of their linearity, include states of superposition. Superpositions are states that consist of a sum over (typically classically distinct) states, such as position, momentum, energy, et cetera. It is expected that a quantum gravity theory must also include superpositions. In the case of a quantum gravity theory that resembles classical general relativity so that the gravitational field remains a manifestation of spacetime curvature, states corresponding to superpositions of spacetime metrics must be considered. Interpreting the meaning of such states, which is an end goal, is an open question, which likely cannot be resolved without recourse to a particular quantum gravity theory \cite{Foo_2021, jensen2011can, giacomini2022quantum, giacomini2020einstein, christodoulou2019possibility, anastopoulos2020quantum, rovelli1991quantum, Wu_2006, crowther2018effective}. This present work considers a superposition of otherwise classical reference frames. The hope of gaining insight into the notion of a superposition of spacetimes or spacetime metrics by first considering a superposition of reference frames on a background spacetime should be apparent.

            The concept and treatment of a superposition of reference frames, both in connection to a quantum gravity theory, and for other purposes, is well-motivated and has been explored in numerous contexts \cite{aharonov1984quantum,  aharonov1967charge, bartlett2007reference, PhysRevA.89.052121, giacomini2019quantum, belenchia2018quantum, de2020quantum, vanrietvelde2020change, palmer2014changing, streiter2021relativistic, greenberger2001inconsistency, mikusch2021transformation, ballesteros2021group, PhysRevD.30.368, angelo2011physics, angelo2012kinematics, pienaar2016relational, smith2016quantum}. In quantum mechanics, as is familiar from classical mechanics, a system under investigation is described from an inertial reference frame. The reference frame is itself an extended system, which permits study of the investigated system from its platform. The reference frame has often been treated as a large (perhaps lab-sized) system, and accordingly, is strictly classical. However, the size of systems that exhibit quantum behavior has been steadily growing with the improvement of experimental technique. It is prudent to inquire about the possibility for a system under investigation to serve as a frame of reference in its own right. A treatment of the problem of a quantum observer making observations of an observer has also been considered in several approaches, including notably Everett's relative state interpretation and relational quantum mechanics \cite{everett1957relative, rovelli1996relational, rovelli2005relational}. This reasoning motivates the incorporation of the observer as part of the quantum system and therefore describes the system from a single Hilbert space. A complementary approach, like the one pursued here, seeks a transformation rule between \emph{coexisting} Hilbert spaces. As will be apparent, this more closely mirrors the state of affairs classically where frames of reference do not play a dynamical role in the theory. In classical theories there exist transformation rules between reference frames for coordinate-dependent quantities and in this approach there is a similar construct. It is also more desirable for the purpose of comparing results in this present work, and in future related work, to the diffeomorphism invariance of General Relativity. To this end, we work in position space. 

In section \ref{Two observers} the case of two reference frames in superposition relative to one another is considered. A complex-valued wavefunctional defined on the set of coordinate transformations between the two reference frames, $\Psi[x(x')]$, is introduced. An example of a superposition of two rotations is considered. In section \ref{Three Observers} a third reference frame is introduced. The closure of coordinate system transformations is extended to reference frames in superposition. The case of a superposition of two rotations is again used as an example. In section \ref{Consistency} a rule for using the wavefunctional to transform wavefunctions is posited and consistency with the Schrodinger equation is demonstrated. 

\section{\label{Two observers}Two Classical Reference Frames in Superposition}

Let O and O' be two frames of reference. Classically, each frame has a set of coordinates (measuring rods) for establishing positions and clocks for recording times. The coordinates will be denoted in the usual way $x_\alpha = (x,y,z,t)$ and $x'_\beta = (x',y',z',t')$. Greek letter indices will run from 1 to 4, while Roman letter indices will run from 1 to 3. Unless otherwise stated, summation notation is in use. When there is no risk of confusion indices may not be written explicitly. There is a transformation between the $O$ and $O'$ frames such that $x_\alpha = x_\alpha(x')$. The transformation must, of course, be invertible so that $x'_\beta= x'_\beta(x)$. 

We wish to consider that these two reference frames are in superposition with respect to one another. What should this mean? A system is in a superposition of ``classical states'' (states with well-defined properties such as position, momentum, energy, et cetera) when a wavefunction is introduced, which assigns a complex number (amplitude) to a set of classical states. For example, a particle is in a superposition of two classical position states, $\mathbf{x}_0$ and $\mathbf{x}_1$, when its wavefunction, $\psi$, is such that \begin{equation}
\psi \propto \delta(\mathbf{x}-\mathbf{x}_0) +\delta(\mathbf{x}-\mathbf{x}_1) . \label{twotermsuperposition}
\end{equation}
To mirror this notion of superposition while extending it to include superposed reference frames, the transformation between the reference frames should not be fixed, but instead replaced with a complex-valued function(al) on the space of transformations. We introduce just such an amplitude $\Psi_{OO'}$ and allow \begin{equation}\label{wavefunctionaldefined1}
    \Psi_{OO'} = \Psi_{OO'} \left[x_\alpha(x')\right].
\end{equation}
Equation \ref{wavefunctionaldefined1} is defined on the ``forward" transformations $x(x')$ (we ignore coordinate indices here for clarity). As will be demonstrated it will take quantities from frame $O'$ to $O$. We expect that there should be a wavefunctional defined on the ``reverse" transformation $x'(x)$ taking quantities from $O$ to $O'$. That is, there must be a wavefunctional $\Psi_{O'O}\left[x'_\alpha(x)\right]$, which is ``inverse" to $\Psi[x_\alpha(x)]$. The precise sense in which these two wavefunctional can be said to be inverse to one another will be made clear in the next section.

This wavefunctional is, by its nature, a relational quantity. We cannot speak of a reference frame in superposition in any absolute sense. We may only state that a reference frame is in superposition relative to another frame. We give a physical interpretation that again mirrors the case of standard quantum mechanics. Two reference frames are in superposition when the the transformation rule between them is completely described by a wavefunctional on the space of transformations $\Psi_{OO'}[x_\alpha(x')]$. If the reference frames were to interact so as to ascertain a specific transformation then the physical interaction would drive the wavefunctional toward a delta functional, which selects one particular transformation. That is to say that if two reference frames were in superposition then a mutual interaction, or coupling, would, in the phrasing of standard quantum mechanics, collapse the superposition. There is an issue with this if the wavefunctional is to be seen as a functional of $x_\alpha(x')$, because $x_\alpha(x')$ contains time as a coordinate and includes the time component of the transformation. If the interaction that led to collapse-like behavior occurred at a time $t_1$ then the transformation function would supposedly be known for all time. If the coordinate systems are, however, again isolated then it is expected that they, by some dynamics, may again evolve to be superposition. This is inconsistent with the claim that the transformation was established at time $t_1$ and remains valid for all times. We remedy this situation in this present work by considering superpositions of time-independent transformations. Then we have, \begin{equation}
   \Psi_{OO'}\left[x_\alpha(x'_\beta)\right]\xrightarrow[]{} \Psi_{OO'}\left[x_i(x'_j)\right],\end{equation} 
   where again Roman indices are restricted to $1,2, \text{and } 3$.
\begin{equation}\Psi_{OO'}[x_i(x'_j)]\xrightarrow[]{interaction} \delta[x_i(x'_j)-x^0_i(x'_j)].\end{equation}
The probability density for the interaction to select $x_i(x'_j)$ is, in accord with the Born rule, \begin{equation}
    \mathcal{P}\left[x_i(x')\right] = \Psi^*_{OO'}\left[x_i(x'_j)\right]\Psi_{OO'}\left[x_i(x'_j)\right]. 
\end{equation} 
The path integral of $\mathcal{P}\left[x_i(x')\right]$ over a subset $S$ of all possible transformations $x_i(x')$ would give the probability of the transformation, upon interaction, to be an element of $S$, in accord with the Born rule \cite{feynman2010quantum}. That is,  
\begin{eqnarray} \nonumber
   &&\int_{S} \mathcal{D}x_1(x')\mathcal{D}x_2(x')\mathcal{D}x_3(x')\mathcal{P}[x_i(x')]=\\ \nonumber &&\text{Probability } O \text{ and } O'' \text{are related by a transformation in } S.
\end{eqnarray}

For clarity it is worthwhile to consider an explicit example. For simplicity assume the transformations between reference frames $O$ and $O'$ are restricted to rotations, so that \begin{equation}x_i(x'_j)=R_{ij}x'^j.\label{rotationonly}
\end{equation}
Upon substituting equation \ref{rotationonly} into equation \ref{wavefunctionaldefined1}, $\Psi_{OO'}\left[R_{ij}x'_j\right] = \Psi_{OO'}\left(R_{ij}\right)$. The wavefunctional is reduced to a function of the parameters of the rotation matrix $R_{ij}$. 
In analogy to the state defined in \ref{twotermsuperposition}, consider a superposition of two possible rotations $R_{ij}(+\theta, \bold{\hat{n}})$ and $R_{ij}(-\theta, \bold{\hat{n}})$ \begin{equation}
    \Psi_{OO'}[x(x')] = \delta\left(R_{ij}-R_{ij}(+\theta, \mathbf{\hat{n}})\right)+\delta\left(R_{ij}-R_{ij}(-\theta, \mathbf{\hat{n}})\right).
\end{equation}
Where delta functions of rotation matrices vanish unless the argument vanishes, thereby enforcing $R_{ij} = R_{ij}(+\theta, \bold{\hat{n}})$ or $R_{ij} = R_{ij}(-\theta, \bold{\hat{n}})$. Note that normalization of the state is ignored here for elegance. 
Upon interaction between the reference frames the transformation between them will be $x_i =R_{ij}(+\theta, \bold{\hat{n}})x'_j $ with 50\% probability and will be $R_{ij}(-\theta, \bold{\hat{n}})x'_j$ the remainder. This example will be further explored in the next section.

\section{\label{Three Observers}Three Observers in Superposition}
When a third reference frame, $O''$, is introduced and the transformation from $O$ to $O'$ as well as the transformation from $O'$ to $O''$ is specified, then the transformation from $O$ to $O''$ is fixed by the action of composition. If $x'(x)$ takes coordinates from $O$ to $O'$ and $x''(x')$ takes coordinates from $O'$ to $O''$, then the composition $x''(x'(x))$ takes coordinates from $O$ to $O''$.
It is expected that if $O$ and $O'$ are in superposition, and the wavefunctional $\Psi_{OO'}[x(x')]$ is known, and that $O'$ and $O''$ are also in superposition with a known wavefunctional $\Psi_{O'O''}[x'(x'')]$, then $O$ and $O''$ are also in superposition and that the wavefunctional $\Psi_{OO''}[x(x'')]$ may be calculated from $\Psi_{OO'}[x(x')]$ and $\Psi_{O'O''}[x'(x'')]$. That is to say, there must be a rule for composing transformations when the systems are in mutual superposition. We do not seek a derivation from first principles but instead, as will be demonstrated, there seems to be only one natural choice.

Imagine that both $\Psi_{OO'}[x(x')]$ and $\Psi_{O'O''}[x'(x'')]$ are proportional to delta functionals that enforce particular transformations between $O$ and $O'$ and $O'$ and $O''$. That is, \begin{eqnarray}\label{singledelta1}\Psi_{OO'}[x(x')]&=& \delta\left[x(x')-x_0(x')\right]\\\label{singledelta2}
\Psi_{O'O''}[x'(x'')]&=& \delta\left[x'(x'')-x'_0(x'')\right]
\end{eqnarray}
The delta functionals enforce $x(x')=x_0(x')$ and $x'(x'')=x'_0(x'')$, where $x_0(x')$ and $x'_0(x'')$, are particular transformation functions. Note that the indices for the coordinates have been suppressed. These wavefunctionals do not represent a superposition. The transformations are fixed, and thus the transformation from $O$ to $O''$ is also fixed. Namely, the transformation from $O$ to $O''$ is formed from composition: $x(x'')=x(x'(x''))=x_0(x'_0(x''))$. As a result it must be that \begin{equation}\label{singledelta}
 \Psi_{OO''}[x(x'')] =  \delta\left[x(x'')-x_0(x'_0(x''))\right]   
\end{equation}
Equation \ref{singledelta} enforces the known transformation from $O$ to $O''$. There is seemingly only one way that equation \ref{singledelta} can be expressed in terms of equations \ref{singledelta1} and \ref{singledelta2}. Multiply equations \ref{singledelta1} and \ref{singledelta2}. The product is \begin{equation}\label{simpleproduct}\delta\left[x(x')-x_0(x')\right]\delta\left[x'(x'')-x'_0(x'')\right].\end{equation} This defines a functional of both $x(x')$ and $x'(x'')$. It should not be worrisome to see the coordinates $x'$ treated as both independent and dependent quantities in the same expression since equation \ref{simpleproduct} being a functional expression depends only on the distinct functions $x(x')$ and $x'(x'')$, while making the physical operation clearer.

We need to construct a functional of a \emph{single} function $x(x'')$. To do so, functionally integrate equation \ref{simpleproduct} over $x(x')$ and $x'(x'')$ with the restriction that $x(x'(x''))=x(x'')$. That is, compute \begin{equation}\label{doubleintegral}\iint_{\mathcal{S}_{x(x'')}}Dx(x')Dx'(x'')\Psi_{OO'}[x(x')]\Psi_{O'O''}[x'(x'')],\end{equation}
where $\mathcal{S}_{x(x'')}$ is the set of paired functions $\left(x(x'),x'(x'')\right)$, such that the composition of $x(x')$ and $x'(x'')$ yields $x(x'')$, that is so that $x(x'(x''))=x(x'')$.  i.e. \begin{equation}\mathcal{S}_{x(x'')}=\{ \left(x(x'), x'(x'')\right)\vert x(x'(x''))=x(x'')\}.\end{equation} The restricted path integral in equation \ref{doubleintegral} yields a functional of a single function $x(x'')$, as a valid transformation composition rule must do. 

Expression \ref{doubleintegral} for wavefunctionals \ref{singledelta1} and \ref{singledelta2} is 
\begin{equation}\label{unevaluatedintegral}\iint_{S_{x(x'')}}Dx(x')Dx'(x'')\delta\left[x(x')-x_0(x')\right]\delta\left[x'(x'')-x'_0(x'')\right].\end{equation} Evaluating functional integrals is generally very difficult. Here we will argue for the result of the integral, and in appendix \ref{DeltaFunctions} a more detailed discussion and conjecture is provided. We claim that equation \ref{unevaluatedintegral} evaluates to \begin{equation}\label{Rightdelta} \delta[x(x'')-x_0(x'_0(x''))],
\end{equation}
which it must if it is to reproduce our expectation.
 Equation \ref{unevaluatedintegral} consists of an integral over $S_{x(x'')}$. The integral sums over a range of functions $x(x')$ and $x'(x'')$. Without restricting the integral to $S_{x(x'')}$, there would be no consistent function $x(x'')$ formed by the composition $x(x'(x''))$). Thus, this restriction ensures that the function $x(x'')$ is well-defined and that \ref{unevaluatedintegral} may be regarded as a functional of $x(x'')$. The integrand in equation \ref{unevaluatedintegral} consists of a product of delta functionals. The first delta functional, and therefore the whole expression, will vanish if $x(x') \neq x_0(x')$. The second functional, and therefore the expression, will vanish if $x'(x'') \neq x_0(x'_0(x''))$. We conclude that expression \ref{unevaluatedintegral} will vanish unless $x(x'')=x_0(x'_0(x''))$. Now consider\begin{widetext}\begin{equation}
    \int Dx(x'')\left( \iint_{S_{x(x'')}}Dx(x')Dx'(x'')\delta\left[x(x')-x_0(x')
    \right]\delta\left[ x'(x'')-x'_0(x'')\right] \right).\end{equation}\end{widetext} 
The double integrals inside the parenthesis are restricted to set $S_{x(x'')}$, while the exterior integral integrates over the restriction. Together, the whole function space is integrated over, and the result is unity.  $\iint_{S_{x(x'')}}Dx(x')Dx'(x'')\delta\left[x(x')-x_0(x')\right]\delta\left[x'(x'')-x'_0(x'')\right] $ vanishes if $x(x'') \neq x_0(x'_0(x''))$ and gives unity when integrated with respect to $x(x'')$. These are the two defining properties of a delta functional in $x(x'')$, which is the claim. 

 For delta functional wavefunctionals equation \ref{doubleintegral} evaluates to equation \ref{Rightdelta}, which is the expected form of the wavefunctional since it enforces the same transformation as equation \ref{singledelta}. We generalize from this result and posit that the composition rule for wavefunctionals is as follows--- \begin{equation}\Psi_{OO''}[x(x'')] = \iint_{S_{x(x'')}}Dx(x') Dx'(x'')\Psi_{OO'}[x(x')]\Psi_{O'O''}[x'(x'')]. \label{therule}\end{equation}

To see how equation \ref{therule} works in the case of a superposition it is worthwhile to return to the case of two rotations in superposition begun in Section \ref{Two observers}. Take $\Psi_{OO'}$ to be
\begin{equation}\label{OO'}
    \Psi_{OO'}(R_{ij}) = A_{OO'}\delta\left(R_{ij}-R_{ij}(+\theta, \bold{\hat{n}})\right)+B_{OO'}\delta\left(R_{ij}-R_{ij}(-\theta, \bold{\hat{n}})\right).
\end{equation}
Where $A_{OO'}$ and $B_{OO'}$ are (possibly complex) coefficients. Similarly, take $\Psi_{O'O''}$ to be \begin{widetext}\begin{equation}\label{O'O''}
    \Psi_{O'O''}(R'_{ij}) = A_{O'O''}\delta\left(R'_{ij}-R_{ij}(+\gamma, \bold{\hat{n'}})\right)+B_{O'O''}\delta\left(R'_{ij}-R_{ij}(-\gamma, \bold{\hat{n'}})\right).
\end{equation}\end{widetext}
The two terms in equation \ref{OO'} indicate that there are two possible transformations in superposition, $x=R(+\theta, \bold{\hat{n}})x'$ or $x=R(-\theta, \bold{\hat{n}})x'$. Likewise from equation \ref{O'O''} there are also two transformations,  $x'=R(+\gamma, \bold{\hat{n'}})x'' $ or $x'=R(-\gamma, \bold{\hat{n'}})x'' $. As a result, there are four transformations between $x$ and $x''$ that are in superposition. They are, $x=R(+\theta, \bold{\hat{n}})R(+\gamma, \bold{\hat{n'}})x''$, $x=R(+\theta, \bold{\hat{n}})R(-\gamma, \bold{\hat{n'}})x''$, $x=R(-\theta, \bold{\hat{n}})R(+\gamma, \bold{\hat{n'}})x''$, or $x=R(-\theta, \bold{\hat{n}})R(-\gamma, \bold{\hat{n'}})x''$. That is, $\Psi_{OO''}$ should have four delta function terms enforcing the transformations. Explicitly, \begin{eqnarray}
    \Psi_{OO''}(R''_{ij}) = \nonumber C\delta\left(R''_{ij}-R_{im}(+\theta, \bold{\hat{n}})R_{mj}(+\gamma, \bold{\hat{n'}})\right) \\ \nonumber + D\delta\left(R''_{ij}-R_{im}(+\theta, \bold{\hat{n}})R_{mj}(-\gamma, \bold{\hat{n'}})\right)\\ \nonumber + E\delta\left(R''_{ij}-R_{im}(-\theta, \bold{\hat{n}})R_{mj}(+\gamma, \bold{\hat{n'}})\right)\\ +F\delta\left(R''_{ij}-R_{im}(-\theta, \bold{\hat{n}})R_{mj}(-\gamma, \bold{\hat{n'}})\right). \label{compareto}
\end{eqnarray}  

Now apply equation \ref{therule}. The product of $\Psi_{OO'}$ and $\Psi_{O'O''}$ is \begin{widetext}\begin{eqnarray} \nonumber &&\Psi_{OO'}\Psi_{O'O''} = \\
    \nonumber &&A_{OO'}A_{O'O''}\delta \left( R_{ij}-R_{ij}(+\theta, \bold{\hat{n}}) \right) \delta\left(R'_{ij}-R_{ij}(+\gamma, \bold{\hat{n'}})\right)+A_{OO'}B_{O'O''}\delta \left( R_{ij}-R_{ij}(+\theta, \bold{\hat{n}}) \right) \delta\left(R'_{ij}-R_{ij}(-\gamma, \bold{\hat{n'}})\right)\\ &+&B_{OO'}A_{O'O''}\delta \left( R_{ij}-R_{ij}(-\theta, \bold{\hat{n}}) \right) \delta\left(R'_{ij}-R_{ij}(+\gamma, \bold{\hat{n'}})\right)+B_{OO'}B_{O'O''}\delta \left( R_{ij}-R_{ij}(-\theta, \bold{\hat{n}}) \right) \delta\left(R'_{ij}-R_{ij}(-\gamma, \bold{\hat{n'}})\right). \label{psiproduct}
\end{eqnarray} \end{widetext}
Following equation \ref{therule} we integrate equation $\Psi_{OO'}\Psi_{O'O''}$ over all functions $x(x')$ and $x'(x'')$, such that their composition $x(x'(x''))$ is $x(x'')$. The transformations are restricted to be rotations only and thus the path integrals become integrals with respect to $R_{ij}$ and $R'_{ij}$. The restriction that $x(x'(x''))$ must equal $x(x'')$ becomes $R''_{ij}=R_{im}R'_{mj}$, where $R''_{ij}$ will represent the transformation from $O$ to $O''$ so that $x_i=R''_{im}x''_m$ if there was no superposition. The four terms of equation \ref{psiproduct} are similar so it is sufficient to consider just one. The integral of the first term in equation \ref{psiproduct} becomes \begin{widetext}\begin{eqnarray}  A_{OO'}A_{O'O''} \iint_{S}dR_{ij}dR'_{qr}&\delta \left( R_{ij}-R_{ij}(+\theta, \bold{\hat{n}})\right)\delta\left(R'_{ij}-R_{ij}(+\gamma, \bold{\hat{n'}})\right) \nonumber =A_{OO'}A_{O'O''}\delta \left( R''_{ij}-R_{im}(+\theta, \bold{\hat{n}})R_{mj}(+\gamma, \bold{\hat{n'}})\right).
\end{eqnarray} \end{widetext}
The three remaining terms are similar. The result is \begin{multline}
  \Psi_{OO''}(R''_{ij})=\\A_{OO'}A_{O'O''}\delta \left( R''_{ij}-R_{im}(+\theta, \bold{\hat{n}})R_{mj}(+\gamma, \bold{\hat{n'}})\right)\\+A_{OO'}B_{O'O''}\delta \left( R''_{ij}-R_{im}(+\theta, \bold{\hat{n}})R_{mj}(-\gamma, \bold{\hat{n'}})\right)\\+B_{OO'}A_{O'O''}\delta \left( R''_{ij}-R_{im}(-\theta, \bold{\hat{n}})R_{mj}(+\gamma, \bold{\hat{n'}})\right)\\ +B_{OO'}B_{O'O''}\delta \left( R''_{ij}-R_{im}(-\theta, \bold{\hat{n}})R_{mj}(-\gamma, \bold{\hat{n'}})\right). \label{theproductsum}
\end{multline}
Comparing equation \ref{theproductsum} with equation \ref{compareto} it is easy to see that the coefficients $C, D, E, \text{and } F$ are the products $A_{OO'}A_{O'O''}$, $A_{OO'}B_{O'O''}$, $B_{OO'}A_{O'O''}$, and $B_{OO'}B_{O'O''}$, respectively. 

Equation \ref{therule} ensures that the composition of any number of wavefunctionals is also a wavefunctional, so that there is closure. Now also consider transforming from the $O$ frame to the $O'$ frame followed by a transformation from the $O'$ frame to the $O$ frame. The total transformation must be the identity transformation, so that $x(x'')=1$, and the wavefunctionals must satisfy an identity. Using equation \ref{therule} we have
\begin{equation}1= \iint_{S_1}Dx(x') Dx'(x'')\Psi_{OO'}[x(x')]\Psi_{O'O}[(x'(x''))], \label{theidentity}\end{equation}
where $S_1$, not to be confused with the unit circle, is the set of transformation functions that compose to give the identity. That is to say, the integral is restricted to a set of functions $x(x')$ and $x'(x'')$ such that $x(x'(x''))=1$. Equation \ref{theidentity} makes precise the notion of an inverse wavefunctional. The similarity between equations \ref{therule} and \ref{theidentity} and matrix multiplication and matrix inverses should be apparent.

\section{\label{Consistency}Consistency with the Schr\"{o}dinger equation}

The Schr\"{o}dinger equation is invariant under rigid Euclidean motions (rotations and translations). Let us prove that the Schr\"{o}dinger equation is invariant under any transformation to a superposition of rotated or translated frames. That is, the present formalism is consistent with Schr\"{o}dinger equation. Assume that relative to a frame $O$ there is a system that is studied, which has a wavefunction $\psi(x,t)$. Now consider a second frame $O'$ in superposition relative $O$. There is a wavefunctional $\Psi[x(x')]$ which governs the transformation between the two superposed frames. How are we to transform the known wavefunction $\psi(x,t)$ in the $O$ frame to the wavefunction that would be associated to it in the $O'$ frame? We hypothesize the following transformation rule: \begin{equation}\psi'(x',t) = \int_S \mathcal{D}x(x') \Psi[x(x')]\psi(x(x'),t). \label{wavefunctiontrans}\end{equation} where $S$ is the set of functions $x(x')$. The transformation in equation \ref{wavefunctiontrans} is natural. There are only a few distinct rules that can transform a function $\psi(x,t)$ to a function of variables $x'$, and it ensures that if the system described by $\psi(x,t)$ happened to be in position eigenstate relative to $O$ then $O'$, assumed to be in a superposition relative to $O$, would associate a wavefunction $\psi'(x,t)$ to the system that corresponds to a superposition. This is ensured because eq. \ref{wavefunctiontrans} obviously superposes $\psi(x(x'),t)$ for each function $x(x')$ for which $\Psi[x(x')]$ does not vanish. 
We can extend equation \ref{wavefunctiontrans} to wavefunctions of multi-particle systems as follows. Let $\psi(x_1,x_2,...,x_n,t)$ be the wavefunction of a multi-particle system in the $O$ frame. A generic transformation from the frame $O'$ to $O$ will consist of map $X: O\rightarrow O'$ that takes as input the coordinates in $O'$ and outputs the coordinates in $O$. Let $x'_1$ map to $x_1$ under $X$, and let $x'_2$ map to $x_2$, and so on. Then the multi-particle extension of equation \ref{wavefunctiontrans} takes the form --- \begin{eqnarray} \nonumber
    &\psi&'(x'_1,x'_2,...,x'_n,t)= \\&\int_S& \mathcal{D}X(x') \Psi[X(x')]\psi(X(x'_1),X(x'_2),..., X(x'_n),t). \label{wtmp}
\end{eqnarray}
Note that the wavefunctional, $\Psi$, and the path integral measure, $\mathcal{D}X$ depend only on the form of the function $X$ and not on the value of the $X$ for any particular input $x'$, despite writing $\Psi[X(x')]$ and $\mathcal{D}X(x')$ in equation \ref{wtmp}. 

Using equations \ref{wavefunctiontrans} and \ref{wtmp}, we can now transform derivatives as well, such as $\frac{\partial \psi'(x',t)}{\partial x^{'i}}$. Take the derivative of eq. \ref{wavefunctiontrans} with respect to $x^{'i}$ --- \begin{equation} \frac{\partial \psi'(x',t)}{\partial x^{'i}} = \int_S \mathcal{D}x(x') \Psi[x(x')]\frac{\partial \psi(x,t)}{\partial x^{j}}\frac{\partial x^j}{\partial x^{'i}}. \label{derivativetransform}
\end{equation}  

To extend equation \ref{derivativetransform} to multiple particles we adopt the following notation, $x_{I}^i$ is the $i^{th}$ spatial component of the $I^{th}$ particle. Consequently, \begin{equation}
   \frac{\partial \psi'}{\partial x_I^{'i}}= \int_S \mathcal{D}X(x') \Psi[X(x')]\frac{\partial \psi}{\partial X_I^{j}}\frac{\partial X^j}{\partial x^{'i}},
\end{equation}
where $\frac{\partial \psi}{\partial X_I^{j}}$ means differentiate $\psi$ with respect to the $j^{th}$ spatial component of the $I^{th}$ particle, while $\frac{\partial X^j}{\partial x^{'i}}$ does not require an $I$ index because $X(x')$ is the same function for every particle. 
The Schrodinger equation is invariant under Euclidean transformations and so we restrict our superposition of coordinate systems to a superposition of Euclidean transformations. As a result, $\frac{\partial \psi}{\partial X_I^{j}}\frac{\partial X^j}{\partial x^{'i}}$ would be a rotation, and we can readily compute the transformation of the Laplacian from equation \ref{wtmp}.
\begin{equation}
   \frac{\partial \psi'}{\partial x_I^{'j}\partial x_I^{'i}}= \int_S \mathcal{D}X(x') \Psi[X(x')]\frac{\partial \psi}{\partial X_I^{n} X_I^{m}}\frac{\partial X^n}{\partial x^{'j}}\frac{\partial X^m}{\partial x^{'i}}.\label{laplacetrans}
\end{equation}
Letting $i=j$, summing over the repeated indices, and noting that $\frac{\partial X^n}{\partial x^{'j}}\frac{\partial X^m}{\partial x^{'i}}=\delta^{nm}$ yields the desired transformation --- \begin{equation}
     \frac{\partial \psi'}{\partial x_I^{'i}\partial x_I^{'i}}= \nabla_I^2 \psi'= \int_S \mathcal{D}X(x') \Psi[X(x')]\nabla_I^2 \psi.
\end{equation}
It is easy to see that \begin{equation}
    \frac{\partial \psi'}{\partial t} = \int_S \mathcal{D}X(x') \Psi[X(x')]\frac{\partial \psi}{\partial t}. \label{tderivativetransform}
\end{equation}
The Schrodinger equation for $N$ particles in the $O$ reference frame is as follows--- \begin{equation}
    -\sum_{I=1}^{N}\frac{\hbar^2}{2m_I}\nabla_I^2\psi+V(x_1,x_2,...,x_{3N})\psi = i\hbar\frac{\partial \psi}{\partial t},\label{SE}
\end{equation}
where each of the particles may have a different mass $m_I$ and $V$ could depend upon any of the $3N$ configuration space coordinates.
We may now transform the Schrodinger equation \ref{SE} to a system in superposition relative to the $O'$ reference frame, which is assumed to be in superposition relative to the $O$ frame, with a wavefunctional $\Psi[X(x')]$. Multiply equation \ref{SE} by $\Psi[X(x')]$, let $x_i=X(x_i')$, and sum over all functions $X(x')$. The result is \begin{eqnarray}  \nonumber &-&\sum_{I=1}^{N}\frac{\hbar^2}{2m_I}\int_S\mathcal{D}X(x')\Psi[X(x')]\nabla_I^2\psi+\\ \nonumber &\int_S& \mathcal{D}X(x')\Psi[X(s')]V(X(x_1'),X(x_2'),...,X(x_{3N}'))\psi \\ &=& i\hbar\int_S \mathcal{D}X(x')\frac{\partial \psi}{\partial t}\end{eqnarray} 
Using equations \ref{laplacetrans} and \ref{tderivativetransform} we have \begin{widetext} \begin{eqnarray}
    -\sum_{I=1}^{N}\frac{\hbar^2}{2m_I}\nabla_I^2\psi'+\int_S \mathcal{D}X(x')\Psi[X(x')]V(X(x'_1),X(x'_2),...,X(x'_{3N}))\psi = i\hbar\frac{\partial \psi'}{\partial t} ,\label{halftransSE}
\end{eqnarray} \end{widetext}
where it should be apparent that $\nabla_I^2$ acting on $\psi'$ is $\sum_i\frac{\partial^2}{\partial x'_i \partial x'_i} $.
The potential energy must be invariant with respect to any Euclidean transformations (generally by being a function of the Euclidean distance of the differences of the coordinates). Therefore, \begin{equation}
    V(X(x_1'),X(x_2'),...,X(x_{3N}'))=V(x_1',x_2',...,x_{3N}'),
\end{equation} 
and \begin{multline}
    \int_S \mathcal{D}X(x')\Psi[X(x')]V(X(x_1'),X(x_2'),...,X(x_{3N}'))\psi=\\\int_S \mathcal{D}X(x')\Psi[X(x')]V(x_1',x_2',...,x_{3N}')\psi \label{Vconstant}.
\end{multline}
In equation \ref{Vconstant}, $V(x_1',x_2',...,x_{3N}')$ is a constant relative to the path integral over $X$, and so comes out to yield 
\begin{eqnarray} \nonumber \int_S \mathcal{D}X(x')\Psi[X(x')]V(X(x_1'),X(x_2'),...,X(x_{3N}'))\psi=\\ \nonumber V(x_1',x_2',...,x_{3N}')\int_S \mathcal{D}X(x')\Psi[X(x')]\psi \\ =  V(x_1',x_2',...,x_{3N}')\psi', \label{VWtransform}
\end{eqnarray}
where we have used equation \ref{wtmp} in the last equality. Substituting equation \ref{VWtransform} into equation \ref{halftransSE} replaces the potential energy term with $V(x_1',x_2',...,x_{3N}')\psi'$, and so demonstrates that under a superposition of transformations \begin{widetext} \begin{multline}-\sum_{I=1}^{N}\frac{\hbar^2}{2m_I}\nabla_I^2\psi+V(x_1,x_2,...,x_{3N})\psi = i\hbar\frac{\partial \psi}{\partial t} \xrightarrow{\text{transformation to superposed coordinates} }\\ -\sum_{I=1}^{N}\frac{\hbar^2}{2m_I}\nabla_I^{2}\psi'+V(x_1',x_2',...,x_{3N}')\psi' = i\hbar\frac{\partial \psi'}{\partial t}, \end{multline} \end{widetext}
which was to be shown.

\section{Discussion and Future Work}
This framework demonstrates the feasibility for coordinate systems to be considered in superposition relative to one another via a (complex-valued) wavefunctional $\Psi[x(x')]$, which takes as input transformation functions. It is posited that interaction between the frames will introduce collapse-like behavior and select a particular transformation amongst the superposed transformations with a probability given by the Born rule. A rule for composing wavefunctionals is provided. It is shown that with this composition rule, equation \ref{therule}, they are closed. The wavefunctional, as introduced here, permits the transformation of wavefunctions ``relative" to one frame to be expressed relative to another. This transformation of wavefunctions is consistent with the standard Schrodinger equation.

It is expected that frames may evolve to form superpositions so that after interaction between the frames, which may act to select a particular transformation $x(x')$ by destroying superposition, is removed, they again will evolve into superposition. That is to say that this framework suggests, but does not single out, a kind of frame dynamics. Because of this possibility only the superposition of time-independent transformations is considered in this present work. The extension of this framework to time-dependent transformations, such as the Galilean transformations or Lorentz transformations and the form of the frame dynamics is a matter of future work. 

\appendix

\section{Integral of Delta Functions over Measure Zero Sets}
\label{DeltaFunctions}
Consider the integral \begin{equation}
 \int_Ldx \delta (x-x_0) =  \left\{
\begin{array}{ll}
      0 & x_0\notin L \\
      1&x_0\in L  \\

\end{array} 
\right.
\end{equation}
Consequently, 
 \begin{equation}
 \frac{1}{L}\int_Ldx \delta (x-x_0) =  \left\{
\begin{array}{ll}
      0 & x_0\notin L \\
      \frac{1}{L}&x_0\in L  \\

\end{array} 
\right. \end{equation}
If the limit as $L$ approaches zero is taken, while $L$ is centered at $X$, then \begin{equation} \lim_{L \xrightarrow{} 0}\int_L dx\delta\left(x-x_0\right) = \delta(X-x_0)
\end{equation}
$L$ is an interval in the limit whose length approaches zero. After the limit, it is a measure zero set, namely that of a single point. 

Consider the integral in two dimensions.
\begin{equation}\int_{A}d^2x\delta(x-x_0)\delta(y-y_0)=  \left\{
\begin{array}{ll}
      0 & (x_0,y_0)\notin A \\
      1&(x_0,y_0)\in A  \\

\end{array} 
\right. \end{equation}
Or for infinitesimal area $A$ centered at $(X,Y)$,
\begin{equation}\label{smallarea}\lim_{A\xrightarrow[]{} 0} \frac{1}{A}\int_{A}d^2x\delta(x-x_0)\delta(y-y_0)=\delta(X-x_0)\delta(Y-y_0)\end{equation} 
The area, however, may be made infinitesimal in many ways. For example, the area may be decreased by choosing points that form a curve. More precisely, let $F(x,y)=0$ be a curve, and let $S$ be a set of points that form the curve defined by $F$ and collection of points that lie infinitesimally around it so that $S$ has area $A$. We conjecture, 
\begin{equation}\label{smallarea}\lim_{A\xrightarrow[]{} 0} \frac{1}{A}\int_{S}d^2x\delta(x-x_0)\delta(y-y_0)=k\delta(F(x_0,y_0)).\end{equation}
Where the limit that $A$ approaches zero is taken such that the remaining elements of $S$ are those points that satisfy $F(x,y)=0$. We have inserted a possible normalization constant $k$, whose exact value would be computed in a proof. More generally, let $S$ be a measure zero set, defined by a membership function $F$, which returns $0$ for elements in the set. That is, \begin{equation}
S = \{(x_1, x_2,..., x_N)\vert F(x_1, x_2,..., x_N)=0\}.    
\end{equation} We conjecture the following,
\begin{multline}\label{Nowvolume}\lim_{V\xrightarrow[]{} 0} \frac{1}{V}\int_{S}d^{N}x\delta(x_1-x^0_1)\delta(x_2-x^0_2)\cdots\delta(x_N-x^0_N)\\=k\delta(F(x^0_1,x^0_2,\ldots,x^0_{N}).\end{multline}
Where the area $A$ is generalized to a volume $V$, and the coordinates are renamed in an obvious way \cite{hormander2015analysis}. 
 From equation \ref{Nowvolume} it follows in $N$ even dimensions that, \begin{eqnarray}
   & \lim_{V\xrightarrow[]{} 0}& \frac{1}{V}\int_{S}d^{\frac{N}{2}}xd^{\frac{N}{2}}y\left[\delta(x_1-x^0_{1})\cdots \delta(x_{\frac{N}{2}}-x^0_{\frac{N}{2}})\right] \nonumber \\ && \nonumber\times \left[\delta(y_1-y^0_{1})\cdots \delta(y_{\frac{N}{2}}-y_{^0\frac{N}{2}})\right] \\ &=& k\delta(F(x^0_1, x^0_2,\ldots x^0_{\frac{N}{2}};y^0_1, y^0_2,\ldots y^0_{\frac{N}{2}}))\label{largefiniteN}.\end{eqnarray}

In the limit that the number of dimensions approaches infinity, we have the \begin{eqnarray}\label{contlimit} \nonumber &\int_S & Dg(x)Dh(x)\delta[g(x)-g_0(x)]\delta[h(x)-h_0(x)] \\ &=& k\delta\left[F[g_0(x),h_0(x)\right]],\end{eqnarray}
where we are integrating over set $S$, a set of paired functions $\left(g(x),h(x)\right)$, defined by a membership function $F\left(g(x),h(x)\right)$. $F\left(g(x),h(x)\right)$ returns $0$ for paired functions $\left(g(x),h(x)\right)$ in $S$. We do not prove that a set of functions defined by a membership function form a zero measure set, which would be difficult. However, in $R^n$ subsets chosen by a membership function are, baring contrivance, zero volume measure sets and we operate under the assumption that this remains valid in the large $N$ limit.   
In equation \ref{unevaluatedintegral}, the membership function is as follows. A pair of functions, $(x(x'),x'(x''))$, are in set $S_{x(x'')}$ if $x(x'(x'')) = x(x'')$. Therefore, $F[x_0(x'),x'_0(x'')] = x(x'')-x_0(x'_0(x''))$ and we have that 
\begin{multline}\int_{S_{x(x'')}} Dx(x')Dx'(x'')\delta[x(x')-x_0(x')]\delta[x'(x'')-x'_0(x'')] =\\ k\delta\left[x(x'')-x_0(x'_0(x'')\right]].\end{multline}
This is the result that is argued for in the text.

\nocite{*}
\bibliography{aipsamp}

\begin{thebibliography}{41}%
\makeatletter
\providecommand \@ifxundefined [1]{%
 \@ifx{#1\undefined}
}%
\providecommand \@ifnum [1]{%
 \ifnum #1\expandafter \@firstoftwo
 \else \expandafter \@secondoftwo
 \fi
}%
\providecommand \@ifx [1]{%
 \ifx #1\expandafter \@firstoftwo
 \else \expandafter \@secondoftwo
 \fi
}%
\providecommand \natexlab [1]{#1}%
\providecommand \enquote  [1]{``#1''}%
\providecommand \bibnamefont  [1]{#1}%
\providecommand \bibfnamefont [1]{#1}%
\providecommand \citenamefont [1]{#1}%
\providecommand \href@noop [0]{\@secondoftwo}%
\providecommand \href [0]{\begingroup \@sanitize@url \@href}%
\providecommand \@href[1]{\@@startlink{#1}\@@href}%
\providecommand \@@href[1]{\endgroup#1\@@endlink}%
\providecommand \@sanitize@url [0]{\catcode `\\12\catcode `\$12\catcode
  `\&12\catcode `\#12\catcode `\^12\catcode `\_12\catcode `\%12\relax}%
\providecommand \@@startlink[1]{}%
\providecommand \@@endlink[0]{}%
\providecommand \url  [0]{\begingroup\@sanitize@url \@url }%
\providecommand \@url [1]{\endgroup\@href {#1}{\urlprefix }}%
\providecommand \urlprefix  [0]{URL }%
\providecommand \Eprint [0]{\href }%
\providecommand \doibase [0]{http://dx.doi.org/}%
\providecommand \selectlanguage [0]{\@gobble}%
\providecommand \bibinfo  [0]{\@secondoftwo}%
\providecommand \bibfield  [0]{\@secondoftwo}%
\providecommand \translation [1]{[#1]}%
\providecommand \BibitemOpen [0]{}%
\providecommand \bibitemStop [0]{}%
\providecommand \bibitemNoStop [0]{.\EOS\space}%
\providecommand \EOS [0]{\spacefactor3000\relax}%
\providecommand \BibitemShut  [1]{\csname bibitem#1\endcsname}%
\let\auto@bib@innerbib\@empty
\bibitem [{\citenamefont {Foo}, \citenamefont {Mann},\ and\ \citenamefont
  {Zych}(2021)}]{Foo_2021}%
  \BibitemOpen
  \bibfield  {author} {\bibinfo {author} {\bibfnamefont {J.}~\bibnamefont
  {Foo}}, \bibinfo {author} {\bibfnamefont {R.~B.}\ \bibnamefont {Mann}}, \
  and\ \bibinfo {author} {\bibfnamefont {M.}~\bibnamefont {Zych}},\ }\bibfield
  {title} {\enquote {\bibinfo {title} {Schrödinger's cat for de sitter
  spacetime},}\ }\href {\doibase 10.1088/1361-6382/abf1c4} {\bibfield
  {journal} {\bibinfo  {journal} {Classical and Quantum Gravity}\ }\textbf
  {\bibinfo {volume} {38}},\ \bibinfo {pages} {115010} (\bibinfo {year}
  {2021})}\BibitemShut {NoStop}%
\bibitem [{\citenamefont {Jensen}(2011)}]{jensen2011can}%
  \BibitemOpen
  \bibfield  {author} {\bibinfo {author} {\bibfnamefont {R.}~\bibnamefont
  {Jensen}},\ }\bibfield  {title} {\enquote {\bibinfo {title} {Can the universe
  be represented by a superposition of spacetime manifolds?}}\ }\href@noop {}
  {\bibfield  {journal} {\bibinfo  {journal} {Physics Procedia}\ }\textbf
  {\bibinfo {volume} {20}},\ \bibinfo {pages} {47--62} (\bibinfo {year}
  {2011})}\BibitemShut {NoStop}%
\bibitem [{\citenamefont {Giacomini}\ and\ \citenamefont
  {Brukner}(2022)}]{giacomini2022quantum}%
  \BibitemOpen
  \bibfield  {author} {\bibinfo {author} {\bibfnamefont {F.}~\bibnamefont
  {Giacomini}}\ and\ \bibinfo {author} {\bibfnamefont {{\v{C}}.}~\bibnamefont
  {Brukner}},\ }\bibfield  {title} {\enquote {\bibinfo {title} {Quantum
  superposition of spacetimes obeys einstein's equivalence principle},}\
  }\href@noop {} {\bibfield  {journal} {\bibinfo  {journal} {AVS Quantum
  Science}\ }\textbf {\bibinfo {volume} {4}},\ \bibinfo {pages} {015601}
  (\bibinfo {year} {2022})}\BibitemShut {NoStop}%
\bibitem [{\citenamefont {Giacomini}\ and\ \citenamefont
  {Brukner}(2020)}]{giacomini2020einstein}%
  \BibitemOpen
  \bibfield  {author} {\bibinfo {author} {\bibfnamefont {F.}~\bibnamefont
  {Giacomini}}\ and\ \bibinfo {author} {\bibfnamefont {{\v{C}}.}~\bibnamefont
  {Brukner}},\ }\bibfield  {title} {\enquote {\bibinfo {title} {Einstein's
  equivalence principle for superpositions of gravitational fields},}\
  }\href@noop {} {\bibfield  {journal} {\bibinfo  {journal} {arXiv preprint
  arXiv:2012.13754}\ } (\bibinfo {year} {2020})}\BibitemShut {NoStop}%
\bibitem [{\citenamefont {Christodoulou}\ and\ \citenamefont
  {Rovelli}(2019)}]{christodoulou2019possibility}%
  \BibitemOpen
  \bibfield  {author} {\bibinfo {author} {\bibfnamefont {M.}~\bibnamefont
  {Christodoulou}}\ and\ \bibinfo {author} {\bibfnamefont {C.}~\bibnamefont
  {Rovelli}},\ }\bibfield  {title} {\enquote {\bibinfo {title} {On the
  possibility of laboratory evidence for quantum superposition of
  geometries},}\ }\href@noop {} {\bibfield  {journal} {\bibinfo  {journal}
  {Physics Letters B}\ }\textbf {\bibinfo {volume} {792}},\ \bibinfo {pages}
  {64--68} (\bibinfo {year} {2019})}\BibitemShut {NoStop}%
\bibitem [{\citenamefont {Anastopoulos}\ and\ \citenamefont
  {Hu}(2020)}]{anastopoulos2020quantum}%
  \BibitemOpen
  \bibfield  {author} {\bibinfo {author} {\bibfnamefont {C.}~\bibnamefont
  {Anastopoulos}}\ and\ \bibinfo {author} {\bibfnamefont {B.-L.}\ \bibnamefont
  {Hu}},\ }\bibfield  {title} {\enquote {\bibinfo {title} {Quantum
  superposition of two gravitational cat states},}\ }\href@noop {} {\bibfield
  {journal} {\bibinfo  {journal} {Classical and Quantum Gravity}\ }\textbf
  {\bibinfo {volume} {37}},\ \bibinfo {pages} {235012} (\bibinfo {year}
  {2020})}\BibitemShut {NoStop}%
\bibitem [{\citenamefont {Rovelli}(1991)}]{rovelli1991quantum}%
  \BibitemOpen
  \bibfield  {author} {\bibinfo {author} {\bibfnamefont {C.}~\bibnamefont
  {Rovelli}},\ }\bibfield  {title} {\enquote {\bibinfo {title} {Quantum
  reference systems},}\ }\href@noop {} {\bibfield  {journal} {\bibinfo
  {journal} {Classical and Quantum Gravity}\ }\textbf {\bibinfo {volume} {8}},\
  \bibinfo {pages} {317} (\bibinfo {year} {1991})}\BibitemShut {NoStop}%
\bibitem [{\citenamefont {Wu}\ and\ \citenamefont {Zhang}(2006)}]{Wu_2006}%
  \BibitemOpen
  \bibfield  {author} {\bibinfo {author} {\bibfnamefont {X.}~\bibnamefont
  {Wu}}\ and\ \bibinfo {author} {\bibfnamefont {H.}~\bibnamefont {Zhang}},\
  }\bibfield  {title} {\enquote {\bibinfo {title} {Chaotic dynamics in a
  superposed weyl spacetime},}\ }\href {\doibase 10.1086/508129} {\bibfield
  {journal} {\bibinfo  {journal} {The Astrophysical Journal}\ }\textbf
  {\bibinfo {volume} {652}},\ \bibinfo {pages} {1466--1474} (\bibinfo {year}
  {2006})}\BibitemShut {NoStop}%
\bibitem [{\citenamefont {Crowther}(2018)}]{crowther2018effective}%
  \BibitemOpen
  \bibfield  {author} {\bibinfo {author} {\bibfnamefont {K.}~\bibnamefont
  {Crowther}},\ }\href@noop {} {\emph {\bibinfo {title} {Effective
  spacetime}}}\ (\bibinfo  {publisher} {Springer},\ \bibinfo {year}
  {2018})\BibitemShut {NoStop}%
\bibitem [{\citenamefont {Aharonov}\ and\ \citenamefont
  {Kaufherr}(1984{\natexlab{a}})}]{aharonov1984quantum}%
  \BibitemOpen
  \bibfield  {author} {\bibinfo {author} {\bibfnamefont {Y.}~\bibnamefont
  {Aharonov}}\ and\ \bibinfo {author} {\bibfnamefont {T.}~\bibnamefont
  {Kaufherr}},\ }\bibfield  {title} {\enquote {\bibinfo {title} {Quantum frames
  of reference},}\ }\href@noop {} {\bibfield  {journal} {\bibinfo  {journal}
  {Physical Review D}\ }\textbf {\bibinfo {volume} {30}},\ \bibinfo {pages}
  {368} (\bibinfo {year} {1984}{\natexlab{a}})}\BibitemShut {NoStop}%
\bibitem [{\citenamefont {Aharonov}\ and\ \citenamefont
  {Susskind}(1967)}]{aharonov1967charge}%
  \BibitemOpen
  \bibfield  {author} {\bibinfo {author} {\bibfnamefont {Y.}~\bibnamefont
  {Aharonov}}\ and\ \bibinfo {author} {\bibfnamefont {L.}~\bibnamefont
  {Susskind}},\ }\bibfield  {title} {\enquote {\bibinfo {title} {Charge
  superselection rule},}\ }\href@noop {} {\bibfield  {journal} {\bibinfo
  {journal} {Physical Review}\ }\textbf {\bibinfo {volume} {155}},\ \bibinfo
  {pages} {1428} (\bibinfo {year} {1967})}\BibitemShut {NoStop}%
\bibitem [{\citenamefont {Bartlett}, \citenamefont {Rudolph},\ and\
  \citenamefont {Spekkens}(2007)}]{bartlett2007reference}%
  \BibitemOpen
  \bibfield  {author} {\bibinfo {author} {\bibfnamefont {S.~D.}\ \bibnamefont
  {Bartlett}}, \bibinfo {author} {\bibfnamefont {T.}~\bibnamefont {Rudolph}}, \
  and\ \bibinfo {author} {\bibfnamefont {R.~W.}\ \bibnamefont {Spekkens}},\
  }\bibfield  {title} {\enquote {\bibinfo {title} {Reference frames,
  superselection rules, and quantum information},}\ }\href@noop {} {\bibfield
  {journal} {\bibinfo  {journal} {Reviews of Modern Physics}\ }\textbf
  {\bibinfo {volume} {79}},\ \bibinfo {pages} {555} (\bibinfo {year}
  {2007})}\BibitemShut {NoStop}%
\bibitem [{\citenamefont {Palmer}, \citenamefont {Girelli},\ and\ \citenamefont
  {Bartlett}(2014{\natexlab{a}})}]{PhysRevA.89.052121}%
  \BibitemOpen
  \bibfield  {author} {\bibinfo {author} {\bibfnamefont {M.~C.}\ \bibnamefont
  {Palmer}}, \bibinfo {author} {\bibfnamefont {F.}~\bibnamefont {Girelli}}, \
  and\ \bibinfo {author} {\bibfnamefont {S.~D.}\ \bibnamefont {Bartlett}},\
  }\bibfield  {title} {\enquote {\bibinfo {title} {Changing quantum reference
  frames},}\ }\href {\doibase 10.1103/PhysRevA.89.052121} {\bibfield  {journal}
  {\bibinfo  {journal} {Phys. Rev. A}\ }\textbf {\bibinfo {volume} {89}},\
  \bibinfo {pages} {052121} (\bibinfo {year} {2014}{\natexlab{a}})}\BibitemShut
  {NoStop}%
\bibitem [{\citenamefont {Giacomini}, \citenamefont {Castro-Ruiz},\ and\
  \citenamefont {Brukner}(2019)}]{giacomini2019quantum}%
  \BibitemOpen
  \bibfield  {author} {\bibinfo {author} {\bibfnamefont {F.}~\bibnamefont
  {Giacomini}}, \bibinfo {author} {\bibfnamefont {E.}~\bibnamefont
  {Castro-Ruiz}}, \ and\ \bibinfo {author} {\bibfnamefont
  {{\v{C}}.}~\bibnamefont {Brukner}},\ }\bibfield  {title} {\enquote {\bibinfo
  {title} {Quantum mechanics and the covariance of physical laws in quantum
  reference frames},}\ }\href@noop {} {\bibfield  {journal} {\bibinfo
  {journal} {Nature communications}\ }\textbf {\bibinfo {volume} {10}},\
  \bibinfo {pages} {1--13} (\bibinfo {year} {2019})}\BibitemShut {NoStop}%
\bibitem [{\citenamefont {Belenchia}\ \emph {et~al.}(2018)\citenamefont
  {Belenchia}, \citenamefont {Wald}, \citenamefont {Giacomini}, \citenamefont
  {Castro-Ruiz}, \citenamefont {Brukner},\ and\ \citenamefont
  {Aspelmeyer}}]{belenchia2018quantum}%
  \BibitemOpen
  \bibfield  {author} {\bibinfo {author} {\bibfnamefont {A.}~\bibnamefont
  {Belenchia}}, \bibinfo {author} {\bibfnamefont {R.~M.}\ \bibnamefont {Wald}},
  \bibinfo {author} {\bibfnamefont {F.}~\bibnamefont {Giacomini}}, \bibinfo
  {author} {\bibfnamefont {E.}~\bibnamefont {Castro-Ruiz}}, \bibinfo {author}
  {\bibfnamefont {{\v{C}}.}~\bibnamefont {Brukner}}, \ and\ \bibinfo {author}
  {\bibfnamefont {M.}~\bibnamefont {Aspelmeyer}},\ }\bibfield  {title}
  {\enquote {\bibinfo {title} {Quantum superposition of massive objects and the
  quantization of gravity},}\ }\href@noop {} {\bibfield  {journal} {\bibinfo
  {journal} {Physical Review D}\ }\textbf {\bibinfo {volume} {98}},\ \bibinfo
  {pages} {126009} (\bibinfo {year} {2018})}\BibitemShut {NoStop}%
\bibitem [{\citenamefont {de~la Hamette}\ and\ \citenamefont
  {Galley}(2020)}]{de2020quantum}%
  \BibitemOpen
  \bibfield  {author} {\bibinfo {author} {\bibfnamefont {A.-C.}\ \bibnamefont
  {de~la Hamette}}\ and\ \bibinfo {author} {\bibfnamefont {T.~D.}\ \bibnamefont
  {Galley}},\ }\bibfield  {title} {\enquote {\bibinfo {title} {Quantum
  reference frames for general symmetry groups},}\ }\href@noop {} {\ \textbf
  {\bibinfo {volume} {4}},\ \bibinfo {pages} {367} (\bibinfo {year}
  {2020})}\BibitemShut {NoStop}%
\bibitem [{\citenamefont {Vanrietvelde}\ \emph {et~al.}(2020)\citenamefont
  {Vanrietvelde}, \citenamefont {H{\"o}hn}, \citenamefont {Giacomini},\ and\
  \citenamefont {Castro-Ruiz}}]{vanrietvelde2020change}%
  \BibitemOpen
  \bibfield  {author} {\bibinfo {author} {\bibfnamefont {A.}~\bibnamefont
  {Vanrietvelde}}, \bibinfo {author} {\bibfnamefont {P.~A.}\ \bibnamefont
  {H{\"o}hn}}, \bibinfo {author} {\bibfnamefont {F.}~\bibnamefont {Giacomini}},
  \ and\ \bibinfo {author} {\bibfnamefont {E.}~\bibnamefont {Castro-Ruiz}},\
  }\bibfield  {title} {\enquote {\bibinfo {title} {A change of perspective:
  switching quantum reference frames via a perspective-neutral framework},}\
  }\href@noop {} {\bibfield  {journal} {\bibinfo  {journal} {Quantum}\ }\textbf
  {\bibinfo {volume} {4}},\ \bibinfo {pages} {225} (\bibinfo {year}
  {2020})}\BibitemShut {NoStop}%
\bibitem [{\citenamefont {Palmer}, \citenamefont {Girelli},\ and\ \citenamefont
  {Bartlett}(2014{\natexlab{b}})}]{palmer2014changing}%
  \BibitemOpen
  \bibfield  {author} {\bibinfo {author} {\bibfnamefont {M.~C.}\ \bibnamefont
  {Palmer}}, \bibinfo {author} {\bibfnamefont {F.}~\bibnamefont {Girelli}}, \
  and\ \bibinfo {author} {\bibfnamefont {S.~D.}\ \bibnamefont {Bartlett}},\
  }\bibfield  {title} {\enquote {\bibinfo {title} {Changing quantum reference
  frames},}\ }\href@noop {} {\bibfield  {journal} {\bibinfo  {journal}
  {Physical Review A}\ }\textbf {\bibinfo {volume} {89}},\ \bibinfo {pages}
  {052121} (\bibinfo {year} {2014}{\natexlab{b}})}\BibitemShut {NoStop}%
\bibitem [{\citenamefont {Streiter}, \citenamefont {Giacomini},\ and\
  \citenamefont {Brukner}(2021)}]{streiter2021relativistic}%
  \BibitemOpen
  \bibfield  {author} {\bibinfo {author} {\bibfnamefont {L.~F.}\ \bibnamefont
  {Streiter}}, \bibinfo {author} {\bibfnamefont {F.}~\bibnamefont {Giacomini}},
  \ and\ \bibinfo {author} {\bibfnamefont {{\v{C}}.}~\bibnamefont {Brukner}},\
  }\bibfield  {title} {\enquote {\bibinfo {title} {Relativistic bell test
  within quantum reference frames},}\ }\href@noop {} {\bibfield  {journal}
  {\bibinfo  {journal} {Physical Review Letters}\ }\textbf {\bibinfo {volume}
  {126}},\ \bibinfo {pages} {230403} (\bibinfo {year} {2021})}\BibitemShut
  {NoStop}%
\bibitem [{\citenamefont {Greenberger}(2001)}]{greenberger2001inconsistency}%
  \BibitemOpen
  \bibfield  {author} {\bibinfo {author} {\bibfnamefont {D.~M.}\ \bibnamefont
  {Greenberger}},\ }\bibfield  {title} {\enquote {\bibinfo {title} {The
  inconsistency of the usual galilean transformation in quantum mechanics and
  how to fix it},}\ }\href@noop {} {\bibfield  {journal} {\bibinfo  {journal}
  {Zeitschrift f{\"u}r Naturforschung A}\ }\textbf {\bibinfo {volume} {56}},\
  \bibinfo {pages} {67--75} (\bibinfo {year} {2001})}\BibitemShut {NoStop}%
\bibitem [{\citenamefont {Mikusch}, \citenamefont {Barbado},\ and\
  \citenamefont {Brukner}(2021)}]{mikusch2021transformation}%
  \BibitemOpen
  \bibfield  {author} {\bibinfo {author} {\bibfnamefont {M.}~\bibnamefont
  {Mikusch}}, \bibinfo {author} {\bibfnamefont {L.~C.}\ \bibnamefont
  {Barbado}}, \ and\ \bibinfo {author} {\bibfnamefont {{\v{C}}.}~\bibnamefont
  {Brukner}},\ }\bibfield  {title} {\enquote {\bibinfo {title} {Transformation
  of spin in quantum reference frames},}\ }\href@noop {} {\bibfield  {journal}
  {\bibinfo  {journal} {Physical Review Research}\ }\textbf {\bibinfo {volume}
  {3}},\ \bibinfo {pages} {043138} (\bibinfo {year} {2021})}\BibitemShut
  {NoStop}%
\bibitem [{\citenamefont {Ballesteros}, \citenamefont {Giacomini},\ and\
  \citenamefont {Gubitosi}(2021)}]{ballesteros2021group}%
  \BibitemOpen
  \bibfield  {author} {\bibinfo {author} {\bibfnamefont {A.}~\bibnamefont
  {Ballesteros}}, \bibinfo {author} {\bibfnamefont {F.}~\bibnamefont
  {Giacomini}}, \ and\ \bibinfo {author} {\bibfnamefont {G.}~\bibnamefont
  {Gubitosi}},\ }\bibfield  {title} {\enquote {\bibinfo {title} {The group
  structure of dynamical transformations between quantum reference frames},}\
  }\href@noop {} {\bibfield  {journal} {\bibinfo  {journal} {Quantum}\ }\textbf
  {\bibinfo {volume} {5}},\ \bibinfo {pages} {470} (\bibinfo {year}
  {2021})}\BibitemShut {NoStop}%
\bibitem [{\citenamefont {Aharonov}\ and\ \citenamefont
  {Kaufherr}(1984{\natexlab{b}})}]{PhysRevD.30.368}%
  \BibitemOpen
  \bibfield  {author} {\bibinfo {author} {\bibfnamefont {Y.}~\bibnamefont
  {Aharonov}}\ and\ \bibinfo {author} {\bibfnamefont {T.}~\bibnamefont
  {Kaufherr}},\ }\bibfield  {title} {\enquote {\bibinfo {title} {Quantum frames
  of reference},}\ }\href {\doibase 10.1103/PhysRevD.30.368} {\bibfield
  {journal} {\bibinfo  {journal} {Phys. Rev. D}\ }\textbf {\bibinfo {volume}
  {30}},\ \bibinfo {pages} {368--385} (\bibinfo {year}
  {1984}{\natexlab{b}})}\BibitemShut {NoStop}%
\bibitem [{\citenamefont {Angelo}\ \emph {et~al.}(2011)\citenamefont {Angelo},
  \citenamefont {Brunner}, \citenamefont {Popescu}, \citenamefont {Short},\
  and\ \citenamefont {Skrzypczyk}}]{angelo2011physics}%
  \BibitemOpen
  \bibfield  {author} {\bibinfo {author} {\bibfnamefont {R.~M.}\ \bibnamefont
  {Angelo}}, \bibinfo {author} {\bibfnamefont {N.}~\bibnamefont {Brunner}},
  \bibinfo {author} {\bibfnamefont {S.}~\bibnamefont {Popescu}}, \bibinfo
  {author} {\bibfnamefont {A.~J.}\ \bibnamefont {Short}}, \ and\ \bibinfo
  {author} {\bibfnamefont {P.}~\bibnamefont {Skrzypczyk}},\ }\bibfield  {title}
  {\enquote {\bibinfo {title} {Physics within a quantum reference frame},}\
  }\href@noop {} {\bibfield  {journal} {\bibinfo  {journal} {Journal of Physics
  A: Mathematical and Theoretical}\ }\textbf {\bibinfo {volume} {44}},\
  \bibinfo {pages} {145304} (\bibinfo {year} {2011})}\BibitemShut {NoStop}%
\bibitem [{\citenamefont {Angelo}\ and\ \citenamefont
  {Ribeiro}(2012)}]{angelo2012kinematics}%
  \BibitemOpen
  \bibfield  {author} {\bibinfo {author} {\bibfnamefont {R.}~\bibnamefont
  {Angelo}}\ and\ \bibinfo {author} {\bibfnamefont {A.}~\bibnamefont
  {Ribeiro}},\ }\bibfield  {title} {\enquote {\bibinfo {title} {Kinematics and
  dynamics in noninertial quantum frames of reference},}\ }\href@noop {}
  {\bibfield  {journal} {\bibinfo  {journal} {Journal of Physics A:
  Mathematical and Theoretical}\ }\textbf {\bibinfo {volume} {45}},\ \bibinfo
  {pages} {465306} (\bibinfo {year} {2012})}\BibitemShut {NoStop}%
\bibitem [{\citenamefont {Pienaar}(2016)}]{pienaar2016relational}%
  \BibitemOpen
  \bibfield  {author} {\bibinfo {author} {\bibfnamefont {J.}~\bibnamefont
  {Pienaar}},\ }\bibfield  {title} {\enquote {\bibinfo {title} {A relational
  approach to quantum reference frames for spins},}\ }\href@noop {} {\bibfield
  {journal} {\bibinfo  {journal} {arXiv preprint arXiv:1601.07320}\ } (\bibinfo
  {year} {2016})}\BibitemShut {NoStop}%
\bibitem [{\citenamefont {Smith}, \citenamefont {Piani},\ and\ \citenamefont
  {Mann}(2016)}]{smith2016quantum}%
  \BibitemOpen
  \bibfield  {author} {\bibinfo {author} {\bibfnamefont {A.~R.}\ \bibnamefont
  {Smith}}, \bibinfo {author} {\bibfnamefont {M.}~\bibnamefont {Piani}}, \ and\
  \bibinfo {author} {\bibfnamefont {R.~B.}\ \bibnamefont {Mann}},\ }\bibfield
  {title} {\enquote {\bibinfo {title} {Quantum reference frames associated with
  noncompact groups: The case of translations and boosts and the role of
  mass},}\ }\href@noop {} {\bibfield  {journal} {\bibinfo  {journal} {Physical
  Review A}\ }\textbf {\bibinfo {volume} {94}},\ \bibinfo {pages} {012333}
  (\bibinfo {year} {2016})}\BibitemShut {NoStop}%
\bibitem [{\citenamefont {Everett~III}(1957)}]{everett1957relative}%
  \BibitemOpen
  \bibfield  {author} {\bibinfo {author} {\bibfnamefont {H.}~\bibnamefont
  {Everett~III}},\ }\bibfield  {title} {\enquote {\bibinfo {title} {``relative
  state" formulation of quantum mechanics},}\ }\href@noop {} {\bibfield
  {journal} {\bibinfo  {journal} {Reviews of modern physics}\ }\textbf
  {\bibinfo {volume} {29}},\ \bibinfo {pages} {454} (\bibinfo {year}
  {1957})}\BibitemShut {NoStop}%
\bibitem [{\citenamefont {Rovelli}(1996)}]{rovelli1996relational}%
  \BibitemOpen
  \bibfield  {author} {\bibinfo {author} {\bibfnamefont {C.}~\bibnamefont
  {Rovelli}},\ }\bibfield  {title} {\enquote {\bibinfo {title} {Relational
  quantum mechanics},}\ }\href@noop {} {\bibfield  {journal} {\bibinfo
  {journal} {International Journal of Theoretical Physics}\ }\textbf {\bibinfo
  {volume} {35}},\ \bibinfo {pages} {1637--1678} (\bibinfo {year}
  {1996})}\BibitemShut {NoStop}%
\bibitem [{\citenamefont {Rovelli}(2005)}]{rovelli2005relational}%
  \BibitemOpen
  \bibfield  {author} {\bibinfo {author} {\bibfnamefont {C.}~\bibnamefont
  {Rovelli}},\ }\bibfield  {title} {\enquote {\bibinfo {title} {Relational
  quantum mechanics},}\ }in\ \href@noop {} {\emph {\bibinfo {booktitle} {Quo
  vadis quantum mechanics?}}}\ (\bibinfo  {publisher} {Springer},\ \bibinfo
  {year} {2005})\ pp.\ \bibinfo {pages} {113--120}\BibitemShut {NoStop}%
\bibitem [{\citenamefont {Feynman}, \citenamefont {Hibbs},\ and\ \citenamefont
  {Styer}(2010)}]{feynman2010quantum}%
  \BibitemOpen
  \bibfield  {author} {\bibinfo {author} {\bibfnamefont {R.~P.}\ \bibnamefont
  {Feynman}}, \bibinfo {author} {\bibfnamefont {A.~R.}\ \bibnamefont {Hibbs}},
  \ and\ \bibinfo {author} {\bibfnamefont {D.~F.}\ \bibnamefont {Styer}},\
  }\href@noop {} {\emph {\bibinfo {title} {Quantum mechanics and path
  integrals}}}\ (\bibinfo  {publisher} {Courier Corporation},\ \bibinfo {year}
  {2010})\BibitemShut {NoStop}%
\bibitem [{\citenamefont {H{\"o}rmander}(2015)}]{hormander2015analysis}%
  \BibitemOpen
  \bibfield  {author} {\bibinfo {author} {\bibfnamefont {L.}~\bibnamefont
  {H{\"o}rmander}},\ }\href@noop {} {\emph {\bibinfo {title} {The analysis of
  linear partial differential operators I: Distribution theory and Fourier
  analysis}}}\ (\bibinfo  {publisher} {Springer},\ \bibinfo {year}
  {2015})\BibitemShut {NoStop}%
\bibitem [{\citenamefont {Bohm}(1952)}]{bohm1952suggested}%
  \BibitemOpen
  \bibfield  {author} {\bibinfo {author} {\bibfnamefont {D.}~\bibnamefont
  {Bohm}},\ }\bibfield  {title} {\enquote {\bibinfo {title} {A suggested
  interpretation of the quantum theory in terms of" hidden" variables. i},}\
  }\href@noop {} {\bibfield  {journal} {\bibinfo  {journal} {Physical review}\
  }\textbf {\bibinfo {volume} {85}},\ \bibinfo {pages} {166} (\bibinfo {year}
  {1952})}\BibitemShut {NoStop}%
\bibitem [{\citenamefont {Everett}(2015)}]{everett2015theory}%
  \BibitemOpen
  \bibfield  {author} {\bibinfo {author} {\bibfnamefont {H.}~\bibnamefont
  {Everett}},\ }\bibfield  {title} {\enquote {\bibinfo {title} {The theory of
  the universal wave function},}\ }in\ \href@noop {} {\emph {\bibinfo
  {booktitle} {The many-worlds interpretation of quantum mechanics}}}\
  (\bibinfo  {publisher} {Princeton University Press},\ \bibinfo {year}
  {2015})\ pp.\ \bibinfo {pages} {1--140}\BibitemShut {NoStop}%
\bibitem [{\citenamefont {Albert}\ and\ \citenamefont
  {Loewer}(1988)}]{albert1988interpreting}%
  \BibitemOpen
  \bibfield  {author} {\bibinfo {author} {\bibfnamefont {D.}~\bibnamefont
  {Albert}}\ and\ \bibinfo {author} {\bibfnamefont {B.}~\bibnamefont
  {Loewer}},\ }\bibfield  {title} {\enquote {\bibinfo {title} {Interpreting the
  many worlds interpretation},}\ }\href@noop {} {\bibfield  {journal} {\bibinfo
   {journal} {Synthese}\ ,\ \bibinfo {pages} {195--213}} (\bibinfo {year}
  {1988})}\BibitemShut {NoStop}%
\bibitem [{\citenamefont {Vaidman}(2012)}]{vaidman2012probability}%
  \BibitemOpen
  \bibfield  {author} {\bibinfo {author} {\bibfnamefont {L.}~\bibnamefont
  {Vaidman}},\ }\bibfield  {title} {\enquote {\bibinfo {title} {Probability in
  the many-worlds interpretation of quantum mechanics},}\ }in\ \href@noop {}
  {\emph {\bibinfo {booktitle} {Probability in physics}}}\ (\bibinfo
  {publisher} {Springer},\ \bibinfo {year} {2012})\ pp.\ \bibinfo {pages}
  {299--311}\BibitemShut {NoStop}%
\bibitem [{\citenamefont {Healey}(1984)}]{healey1984many}%
  \BibitemOpen
  \bibfield  {author} {\bibinfo {author} {\bibfnamefont {R.~A.}\ \bibnamefont
  {Healey}},\ }\bibfield  {title} {\enquote {\bibinfo {title} {How many
  worlds?}}\ }\href@noop {} {\bibfield  {journal} {\bibinfo  {journal} {Nous}\
  ,\ \bibinfo {pages} {591--616}} (\bibinfo {year} {1984})}\BibitemShut
  {NoStop}%
\bibitem [{\citenamefont {{Born}}(1926)}]{1926ZPhysBorn}%
  \BibitemOpen
  \bibfield  {author} {\bibinfo {author} {\bibfnamefont {M.}~\bibnamefont
  {{Born}}},\ }\bibfield  {title} {\enquote {\bibinfo {title} {{Zur
  Quantenmechanik der Sto{\ss}vorg{\"a}nge}},}\ }\href {\doibase
  10.1007/BF01397477} {\bibfield  {journal} {\bibinfo  {journal} {Zeitschrift
  fur Physik}\ }\textbf {\bibinfo {volume} {37}},\ \bibinfo {pages} {863--867}
  (\bibinfo {year} {1926})}\BibitemShut {NoStop}%
\bibitem [{\citenamefont
  {Schr{\"o}dinger}(1935)}]{schrodinger1935gegenwartige}%
  \BibitemOpen
  \bibfield  {author} {\bibinfo {author} {\bibfnamefont {E.}~\bibnamefont
  {Schr{\"o}dinger}},\ }\bibfield  {title} {\enquote {\bibinfo {title} {Die
  gegenw{\"a}rtige situation in der quantenmechanik},}\ }\href@noop {}
  {\bibfield  {journal} {\bibinfo  {journal} {Naturwissenschaften}\ }\textbf
  {\bibinfo {volume} {23}},\ \bibinfo {pages} {844--849} (\bibinfo {year}
  {1935})}\BibitemShut {NoStop}%
\bibitem [{\citenamefont {Sommerfeld}\ and\ \citenamefont
  {Heisenberg}(1922)}]{sommerfeld1922intensitat}%
  \BibitemOpen
  \bibfield  {author} {\bibinfo {author} {\bibfnamefont {A.}~\bibnamefont
  {Sommerfeld}}\ and\ \bibinfo {author} {\bibfnamefont {W.}~\bibnamefont
  {Heisenberg}},\ }\bibfield  {title} {\enquote {\bibinfo {title} {Die
  intensit{\"a}t der mehrfachlinien und ihrer zeemankomponenten},}\ }\href@noop
  {} {\bibfield  {journal} {\bibinfo  {journal} {Zeitschrift f{\"u}r Physik}\
  }\textbf {\bibinfo {volume} {11}},\ \bibinfo {pages} {131--154} (\bibinfo
  {year} {1922})}\BibitemShut {NoStop}%
\bibitem [{\citenamefont {Schr{\"o}dinger}(2003)}]{schrodinger2003collected}%
  \BibitemOpen
  \bibfield  {author} {\bibinfo {author} {\bibfnamefont {E.}~\bibnamefont
  {Schr{\"o}dinger}},\ }\href@noop {} {\emph {\bibinfo {title} {Collected
  papers on wave mechanics}}},\ Vol.\ \bibinfo {volume} {302}\ (\bibinfo
  {publisher} {American Mathematical Soc.},\ \bibinfo {year}
  {2003})\BibitemShut {NoStop}%
\end{thebibliography}%

\end{document}